\documentclass[12pt]{article}

\usepackage{scicite}

\usepackage{times}
\usepackage{graphicx}% Include figure files
\usepackage{dcolumn}% Align table columns on decimal point
\usepackage{bm}% bold math
\usepackage{hyperref}% add hypertext capabilities
\usepackage{color}
\usepackage{tabularx}
\usepackage{float}
\usepackage{upgreek}
\usepackage{nicefrac} 
\usepackage{enumitem}
\usepackage{placeins}
\usepackage{tabularx}
\usepackage{braket}
\usepackage{acronym}
\usepackage{siunitx}
\usepackage{tikz}
\usepackage{physics}
\usepackage{titling}
\usepackage[capitalise]{cleveref}

\newenvironment{sciabstract}{%
\begin{quote} \bf}
{\end{quote}}
\newcommand{\TwoEightSi}{\ensuremath{^{28}\text{Si}}}
\newcommand{\TwoNineSi}{\ensuremath{^{29}\text{Si}}}
\newcommand{\natSi}{\ensuremath{^{\text{nat}}\text{Si}}}

\newcommand{\comment}[1]{}
\newcommand{\qo}[1]{\ensuremath{\hat{#1}}} 
\newcommand{\dg}[1]{ #1^{\dagger} } 
\newcommand{\qsigma}{\qo{\sigma}}

\title{Optical observation of single spins in silicon} 

\author
{A. T. K. Kurkjian$^{1\ast}$, D. B. Higginbottom$^{1\ast}$, C. Chartrand$^{1}$,\\
E. R. MacQuarrie$^{1, 2}$, J. R. Klein$^{1}$, N. R. Lee-Hone$^{1}$, J. Stacho$^{1}$, \\
C. Bowness$^{1}$, L. Bergeron$^{1}$, A. DeAbreu$^{1}$, N. A. Brunelle$^{1}$, S. R. Harrigan$^{1}$, \\
J. Kanaganayagam$^{1}$, M. Kazemi$^{1}$, 
D. W. Marsden$^{1}$,  T. S. Richards$^{1}$, L. A. Stott$^{1}$, \\ S. Roorda$^{3}$, K. J. Morse$^{1, 2}$, M. L. W. Thewalt$^{1}$, S. Simmons$^{1,2\dagger}$ \\
\\
\normalsize{$^{1}$Department of Physics, Simon Fraser University,}\\
\normalsize{Burnaby, BC V5A 1S6, Canada}\\
\normalsize{$^{2}$Photonic  Inc.,  Vancouver,  BC,  Canada}\\
\normalsize{$^{3}$Département  de physique,  Université  de  Montréal,}\\
\normalsize{Montréal,  QC H3C 3J7,  Canada}\\
\normalsize{$^\ast$These authors contributed equally to this work.}\\
\\
\normalsize{$^\dagger$Corresponding author. Email: s.simmons@sfu.ca.}
}

\date{}

\begin{document}
% \baselineskip24pt
\maketitle

\begin{sciabstract}
    The global quantum internet will require long-lived, telecommunications band photon-matter interfaces manufactured at scale. Preliminary quantum networks based upon photon-matter interfaces which meet a subset of these demands are encouraging efforts to identify new high-performance alternatives. Silicon is an ideal host for commercial-scale solid-state quantum technologies. It is already an advanced platform within the global integrated photonics and microelectronics industries, as well as host to record-setting long-lived spin qubits. Despite the overwhelming potential of the silicon quantum platform, the optical detection of individually addressable photon-spin interfaces in silicon has remained elusive. In this work we produce tens of thousands of individually addressable `$T$ centre' photon-spin qubits in integrated silicon photonic structures, and characterize their spin-dependent telecommunications-band optical transitions. These results unlock immediate opportunities to construct silicon-integrated, telecommunications-band quantum information networks.
\end{sciabstract}

A pinnacle achievement of modern quantum science has been to isolate, control, and harness individual quantum particles such as single charges, single photons, and single spins. Individual atomic centres in solids have been the basis of a wide variety of scientific breakthroughs, including entanglement generation \cite{Delteil2017,Pompili2021}, long-distance teleportation \cite{Pfaff2014}, a loophole-free Bell’s inequality test \cite{hensen2015}, and memory-enhanced quantum communications \cite{Bhaskar2020}. Central to each of these achievements were `photon-spin centres’: solid-state centres which possess spin(s) as well as spin-dependent optical transitions. Such spins may be entangled with one another remotely via photons to form quantum computing and communication networks \cite{Pompili2021}. Many more breakthroughs await photon-spin networks at scale.

A compelling approach to generating large entangled networks of photon-spin centres is via integrated photonics. Unfortunately, the host materials of the most developed integrated photonics platforms, such as Si and InP, are not host materials to the most well-studied photon-spin centres, such as centres in YSO\cite{Raha2020}, YVO \cite{Zhong2018a} and most notably, diamond \cite{Gruber1997,Doherty2013,hensen2015,Pompili2021}. Efforts are underway to engineer integrated diamond photonics at scale \cite{Wan2020}, and promising centres in hosts such as SiC are being developed \cite{Wolfowicz2020,Falk2013} in tandem with the development of SiC integrated photonics \cite{Lukin2020}. In contrast, the silicon integrated quantum photonics platform boasts decades of maturity, including best-in-class integrated single photon detectors \cite{Akhlaghi2015}, a vast library of existing quantum optics components \cite{Thomson2016}, and the ability to directly leverage the global semiconductor microelectronics industry.

The appeal of photon-spin centres in silicon photonic networks has spurred the development of hybrid platforms combining non-silicon photon-spin host materials with silicon photonics \cite{Dibos2018}. However, some of the longest spin coherence times ever measured have been in the silicon host itself \cite{Tyryshkin2012,Saeedi2013}, which can be isotopically purified to remove the magnetic noise due to the spin $1/2$ \TwoNineSi{} nuclear spins. The identification of suitable individually addressable photon-spin centres in silicon could obviate the need to duplicate the development of integrated photonics in an entirely new, and more challenging, material platform.

Despite this opportunity, single spins had not been detected optically in silicon prior to this work. Single spins without optical access have been detected electrically in silicon \cite{Morello2010,pla2012,Maune2012,Buch2013,Veldhorst2014,Kawakami2014,Mi2018,Crippa2019}, single spins have been driven optically and detected electrically \cite{Yin2013}, and single optical centres without electron spins have been recently observed \cite{Redjem2020}. Yet the key goal of imaging individual spins---a breakthrough in 1997 \cite{Gruber1997} which now represents a routine first step for many researchers in the diamond-based photon-spin defect community---has remained elusive for silicon. A central reason for this has been, until recently, the lack of a suitable candidate silicon photon-spin defect to image. Silicon’s high index of refraction---the very property which gives silicon integrated photonics excellent optical mode confinement---means that any candidate photon-spin defect must be comparatively bright to be detected optically. Unfortunately, the photon-spin centres in silicon most widely studied to date, including Phosphorus \cite{Kane1998}, Selenium \cite{Deabreu2019} and Erbium \cite{Kenyon2005,Yin2013} do not emit brightly. Very recently, the $T$ centre \cite{Bergeron2020,MacQuarrie2021} was identified as a photon-spin interface in silicon combining long-lived electron ($>\SI{2}{ms}$) and nuclear ($>\SI{1.1}{s}$) spins and sharp, spin-dependent telecommunications-band optical transitions. This combination of attributes is exceedingly rare among all photon-spin centres across all solid-state hosts that have been studied to date \cite{Zhang2020}.

In this work we produce tens of thousands of individually addressable `micropucks' in commercial silicon-on-insulator integrated photonic wafer material and, using cryogenic confocal microscopy, confirm that each of the micropucks measured has a small number of individually addressable $T$ centres. We resolve the spin-dependent transitions of individual spins for a subset of these $T$ centres subject to a small magnetic field. This represents the first optical identification of single spins in silicon, and directly paves the way for the development of telecommunications-band, silicon-integrated global quantum technology networks.

\paragraph*{The $T$ centre.}

\begin{figure}[!ht]
    \centering
    \includegraphics[width=12cm]{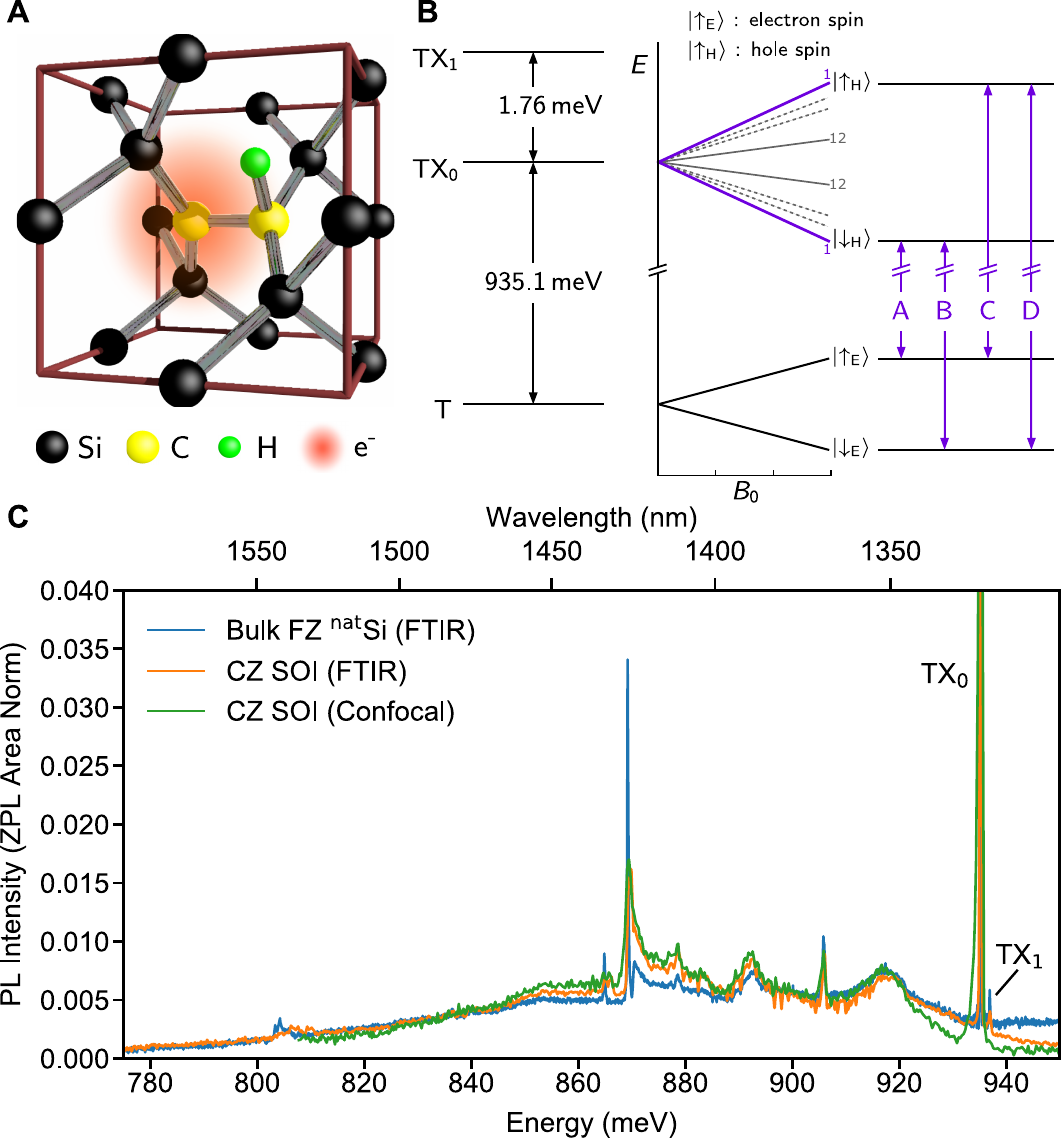}
    \caption{(\textbf{A}) $T$ centre chemical structure \cite{Bergeron2020}. (\textbf{B}) $T$ centre electronic level structure. The two lowest energy bound exciton states are TX$_0$ and TX$_1$. In a magnetic field there are four spin-dependent optical transitions labelled $A,B,C,D$ between the ground state unpaired electron and the unpaired hole spin states of the excited TX$_0$ level for each of the 12 orientational subsets. (\textbf{C}) Photoluminescence (PL) spectra of $T$ centres in bulk natural silicon (blue), SOI wafer (orange), and patterned SOI measured confocally (green, instrument resolution limited). Spectra are normalized to ZPL area.}
    \label{fig:t_review}
\end{figure}

The $T$ centre \cite{Safonov1995, Bergeron2020} is a radiation damage centre in silicon comprising two carbon atoms, one hydrogen atom and an unpaired electron. Its atomic arrangement is illustrated in \cref{fig:t_review}A.
With a zero-phonon-line (ZPL) optical dipole transition at \SI{935.1}{meV} (\SI{1326}{nm}), the $T$ centre is one of a class of radiation damage centres that are known to emit light in the near-infrared telecommunication bands \cite{Chartrand2018}. Measurements of $T$ centre ensembles in isotopically enriched \TwoEightSi{} found an excited state lifetime of \SI{940}{ns}, and ensemble optical transition linewidths as low as \SI{33}{MHz}~\cite{Bergeron2020}.

The $T$ centre ground state features an unpaired electron spin, a hyperfine-coupled hydrogen nuclear spin, as well as up to two additional carbon nuclear spins determined by the choice of carbon isotope constituents. Both the ground state electron spin and the hyperfine-coupled ground state hydrogen nuclear spins are long-lived, having demonstrated electron spin coherence times greater than \SI{2.1}{ms}, and hydrogen nuclear spin lifetimes in excess of \SI{1.1}{s} in ensembles in \TwoEightSi\cite{Bergeron2020}.

The spin and optical level structure of the $T$ centre is shown in \cref{fig:t_review}B. The optically excited state is a bound exciton (BE) in which an electron-hole pair binds to the centre. Within the BE state the two electrons form a singlet, and the unpaired anisotropic spin $3/2$ hole determines the magnetic level structure. The reduced symmetry of the defect splits the hole states into two spin doublets, labeled TX$_0$ and TX$_1$. The signature of these spins is that under a static magnetic field the optical transitions between the two electron spin ground states and the two TX$_0$ hole spin excited states are split into four resolvable spin-dependent transitions. Each $T$ centre belongs to one of twelve orientational subsets relative to a chosen magnetic field direction, each with a possibly different effective hole spin Land\'{e} factor $g_\mathrm{H}$, as shown in \cref{fig:t_review}B that determines the level splittings. It is known that the anisotropic hole $g$ tensor produces effective $g_\mathrm{H}$ values which vary between $0.85$ and $3.50$ depending on the orientation of the centre and the magnetic field \cite{Bergeron2020,Safonov1995}.

\paragraph*{SOI Material.}

We begin by generating high densities of $T$ centres in industry-standard integrated photonic \SI{220}{nm} Czochralski (CZ) \natSi{} device layer silicon-on-insulator (SOI) wafers using an implantation and annealing recipe detailed in Ref. \cite{MacQuarrie2021} and described in the Supplementary Materials. The wafer was subsequently patterned with photonic `micropucks' as well as larger blocks (up to $\SI{200}{\um}\times \SI{200}{\um}$) suitable for optical characterization, whose details are presented later in this manuscript.

The SOI device layer of this sample has a sufficiently high concentration of $T$ centres that $T$ centre photoluminescence (PL) was detectable by Fourier Transform Infrared (FTIR) spectroscopy both prior to and following the patterning of the device layer\cite{Kurkjian2021Methods}. As shown in \cref{fig:t_review}C, the photoluminescence of $T$ centres dominates other fluorescent centres in these spectra and the structure of the phonon sideband (PSB) is visible. Spectra taken before and after patterning reveal an  unchanged, inhomogeneously-broadened ZPL linewidth of \SI{37}{GHz} ---although we will show that the inhomogeneous distribution of centres in the micropuck devices is larger.

In order to optically resolve individual $T$ centres, experiments are conducted upon this sample within a home-built \SI{2.7}{K} cryogenic confocal microscope. We direct PL from a large block SOI section into a single-photon spectrometer \cite{Kurkjian2021Methods} and estimate the temperature of the sample \emph{in situ} from this confocal spectrum. The integrated luminescence ratio of the ZPLs arising from the thermally populated TX$_0$ and TX$_1$ levels, as shown in \cref{fig:t_review}C, reveals a temperature of \SI{4.3(3)}{K} at the SOI device layer \cite{Irion1985a}.

The homogeneous optical linewidth of the ZPL from an individual $T$ centre has a temperature-dependent lower bound, given by the thermal activation between levels TX$_0$ and TX$_1$ \cite{Bergeron2020}. Although this thermal broadening is negligible at \SI{1.5}{K}, it sets an expected lower bound of \SI{255}{MHz} on the observable linewidths of individual emitters at \SI{4.3}{K}. In addition to any power broadening arising from strong resonant optical driving, it is also suspected that a substantial number of individual $T$ centres in this material will be broadened on the order of \SI{1}{GHz} by spectral diffusion \cite{MacQuarrie2021}. The large inhomogeneous broadening of the $T$ centre ZPL observed in this material, over \SI{37}{GHz}, indicates that individual $T$ centres with linewidths on the order of \SI{1}{GHz} could be spectrally addressable if sufficiently low concentrations can be isolated spatially.

To achieve \SI{<1}{GHz} spectral resolution, far below the $\sim$\SIrange{100}{200}{GHz} resolution of our grating spectrometer, resonant excitation methods are required. In this work we resonantly excite single $T$ centres with a tunable single-frequency laser and collect photons emitted into the PSB, a technique known as photoluminescence excitation (PLE) \cite{Kurkjian2021Methods}.

\paragraph*{Silicon micropucks.}\label{sec:pucks}

SOI is a leading photonics platform in part because the high refractive index of silicon ($n \approx 3.5$ at \SI{1326}{nm}) efficiently traps light within the device layer by total internal reflection. This poses a challenge for microscopy on unpatterned SOI, where extraction efficiencies from above the device layer are typically on the order of $0.5$\%. Yet, bright individual $G$ centres have been measured in Si by confocal microscopy \cite{Redjem2020}, with up to \SI{16}{kcps} fluorescence when saturated with above-bandgap excitation \cite{Durand2020}. For longer-lived centres such as $T$ centres it is advantageous to improve the collected luminescence through photonic design. Luminescence from a solid state emitter may be improved using nanophotonic structures, such as planar bullseye gratings \cite{Li2015a} and vertical pillar cavities\cite{Ding2016}.

\begin{figure*}[!ht]
    \centering
    \includegraphics[width=12cm]{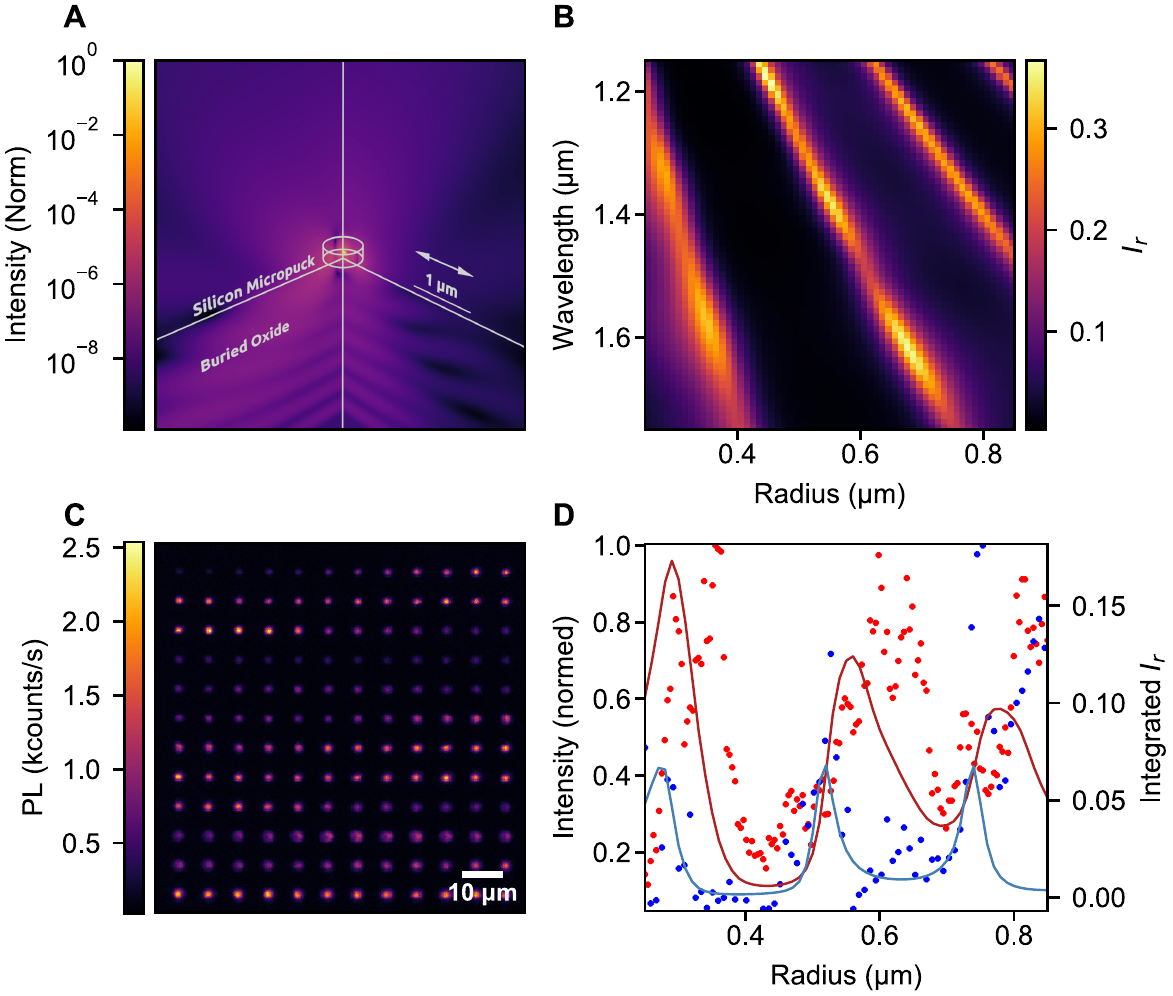}
    \caption{(\textbf{A}) Simulated emission profile of a dipole emitter oriented parallel to the right plane as indicated by the arrow at the centre of a \SI{305}{nm} radius micropuck. Normalized emission intensity (colour axis) is shown on two perpendicular cross-sectional planes. (\textbf{B}) Simulated relative intensity (colour axis) collected by an NA~$=0.7$ microscope objective from a planar emitter at the micropuck centre as a function of wavelength and micropuck radius. (\textbf{C}) Confocal microscope image of micropucks, incrementing in radius from \SI{250}{nm} (upper left) to \SI{850}{nm} (bottom right), by integrating PL signal over $\lambda > \SI{1.33}{\um}$. (\textbf{D}) Simulated (solid) vs measured (dots) intensity of the $T$ ZPL area (blue) and the integrated sideband intensity (red) in PL as a function of micropuck radius. Simulated data is plotted as relative intensity into the objective (right axis) and measured data is peak value normalised (left axis).
        \label{fig:upucks}
    }
\end{figure*}

We take a similar approach to increase collected PL from $T$ centres by fabricating SOI micropucks. These micropucks consist of a silicon cylinder of variable radius $r$ with a thickness determined by the \SI{220}{nm} silicon device layer. Compared to unpatterned SOI, micropucks can increase the outcoupling of light from an emitter by Purcell enhancement of the radiative emission rate as well as improved collection efficiency by shaping the spatial profile of the emitted light. \cref{fig:upucks}A shows the field distribution of a dipole emitter at the micropuck centre, oriented in the device plane, simulated using Lumerical FDTD.

The radius of a micropuck determines the wavelength-dependent ``relative intensity'' of any emitters contained within it. Here the relative intensity ($I_\textrm{r}(\lambda)$) of the emitter compared to a homogeneous silicon environment is given by the product of its collection efficiency ($\eta_\textrm{obj}(\lambda)$) and its Purcell factor ($P_\textrm{F}(\lambda)$). \cref{fig:upucks}B shows the relative intensity collected into an NA~$=0.7$ objective as a function of micropuck radius $r$ according to simulations. Similar simulations reveal a modest single-wavelength Purcell enhancement with a range of \numrange{0.8}{3.5}, approximately periodic with $2 n r /\lambda$. With this information we determine the optimal micropuck radii for PL and PLE measurements of individual optimally positioned $T$ centres.

The solid blue line in \cref{fig:upucks}D is the collected ZPL intensity in PL according to simulations. It is $I_\mathrm{r}$ integrated over the bulk $T$ centre ZPL or, equivalently, the product of $I_\mathrm{r}(\lambda=1326$~nm$)$ and the $T$ centre Debye-Waller factor $0.23$ \cite{Bergeron2020}. From this calculation we estimate a 58-fold ZPL intensity improvement between a centered planar $T$ centre in a \SI{520}{nm} radius micropuck compared to a similarly positioned $T$ centre in unpatterned SOI.  The solid red line in \cref{fig:upucks}D is the collected $T$ centre PSB intensity, which we arrive at by integrating $I_\mathrm{r}(\lambda)$ over the bulk $T$ centre PSB spectrum \cite{Bergeron2020}, specifically \SIrange{1330}{1600}{nm}.

We confirm these predictions experimentally. \cref{fig:upucks}C shows a rastered confocal microscope PL image of a block of micropucks with radii that increment from \SI{250}{nm} (upper left) to \SI{850}{nm} (lower right). The micropucks are excited with \SI{978}{nm} above-bandgap light and the detected fluorescence is filtered to detect wavelengths longer than \SI{1330}{nm} \cite{Kurkjian2021Methods}. As expected, there is a clear periodic variation in integrated intensity with radius, as well as a steady increase in intensity with radius corresponding to the increasing number of $T$ centres in larger micropucks. We find good agreement between the PSB micropuck intensity (red points, \cref{fig:upucks}D) and the simulated PL intensity (red curve, \cref{fig:upucks}D), even though multiple fluorescent centres, including $G$ centres and multiple $T$ centres, contribute to the measured signal at these wavelengths using PL.

We also find good agreement with the collected $T$ centre ZPL intensity in PL as a function of micropuck radius (blue points and blue line, \cref{fig:upucks}D). We record PL spectra from each micropuck confocally \cite{Kurkjian2021Methods}. Typical PL spectra for micropucks with a selection of radii are shown in the Supplementary Material. The blue points \cref{fig:upucks}D are the area of a Gaussian-Lorentzian product fit to each ZPL. Clear $T$ ZPLs on the micropuck spectra confirm the presence of $T$, encouraging the use of PLE techniques to spectrally resolve individual centres.

\paragraph*{Resonant excitation of single centres.}\label{puckPLE}

We next perform confocal PLE on a characteristic area within a large block device. The microscope addresses a near diffraction-limited spot in the silicon device layer. In contrast to $G$ centres which are known to `bleach' under resonant excitation and require co-excitation by both a resonant and above-bandgap laser \cite{Chartrand2018,Redjem2020}, we record continuous $T$ centre fluorescence with only resonant excitation. We find an inhomogeneously broadened $T$ centre ensemble with a linewidth of \SI{30}{GHz} in this spot, which is somewhat narrower than the \SI{37}{GHz} ZPL linewidth observed from the larger SOI sections in PL both before and after device patterning. As shown in Supplementary Materials \cref{fig:soiple}, the structure of the PLE spectrum is typical of ensembles on the verge of single centre resolution.

\begin{figure*}[!ht]
    \centering
    \includegraphics[width=12cm]{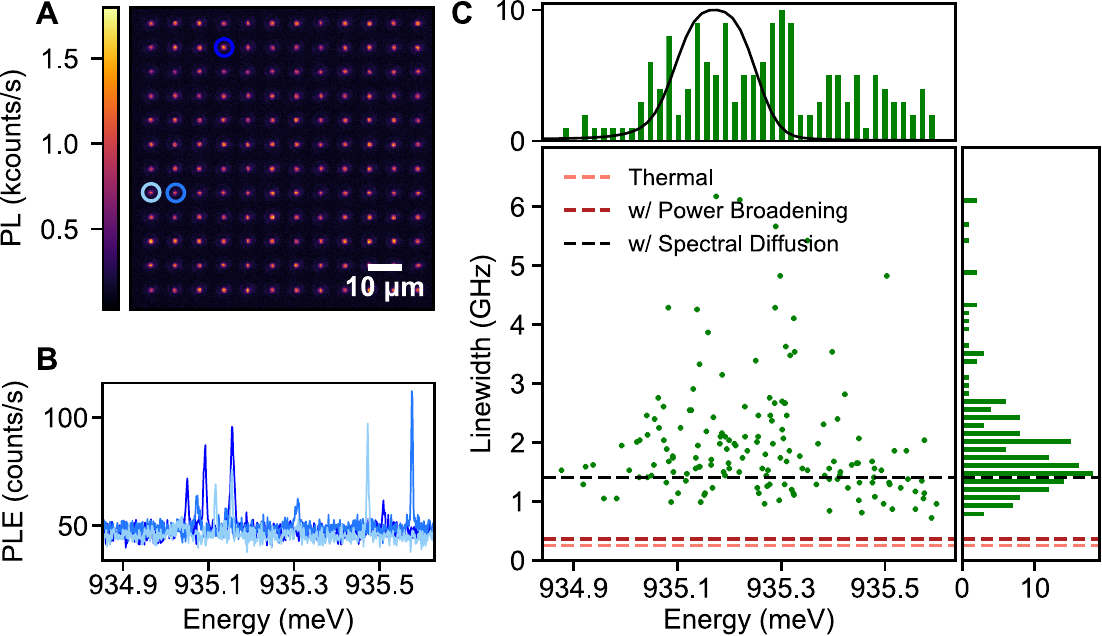}
    \caption{(\textbf{A}) PL raster scan of a block of \SI{305}{nm} radius micropucks. Fluorescence is filtered for $\lambda > \SI{1.33}{\um}$. (\textbf{B}) Characteristic PLE spectra of individual micropucks corresponding in colour to the micropucks circled in A. In these measurements the detected $T$ centre phonon sideband fluorescence is filtered for $\lambda > \SI{1.35}{\um}$. (\textbf{C}) Scatterplot and histograms of all observed $T$ ZPL positions (top) and linewidths (right) from all $144$ micropucks shown in A. The PL spectrum of the SOI material prior to patterning shown in \cref{fig:t_review}C is included for comparison (top, black line).
        \label{fig:ple}
    }
\end{figure*}

Confocal PLE of micropucks reveals the first evidence of addressable single $T$ centres. Based upon the results of the previous section, we select a set of \SI{305}{nm} radius micropucks to image with PL (\cref{fig:ple}A) and then, for each micropuck, measure PLE spectra over a \SI{1.1}{nm} (\SI{776}{\micro\eV}) spectral range around the bulk TX$_0$ ZPL wavelength. Three example single-puck PLE spectra are shown in \cref{fig:ple}B. Each PLE spectrum from this set of $144$ micropucks contains a small number (on average 1.1) of narrow, spectrally isolated resonances sampled from the much larger inhomogeneous $T$ centre distribution. The sparse and resolvable $T$ centres in each micropuck, some with linewidths over 40 times narrower than the inhomogeneously broadened material ensemble, provide strong evidence that this sample offers tens of thousands of individually addressable single $T$ centres for study. From the occurrence of discrete single $T$ PLE resonances we infer a lower bound $T$ centre concentration of $1.7\times 10^{13}$~cm$^{-3}$.

We detect up to $70$~cps of sideband fluorescence from our brightest micropuck-coupled $T$ centres, or $2.1$~kcps after accounting for measured detection losses. This compares favourably to fluorescence rates from prominent telecommunications-band photon-spin centres in Si. Er$^{3+}$, for example, fluoresces at only $500$~cps assuming perfect coupling and no losses at all \cite{Kenyon2005}. From the bulk silicon $T$ centre lifetime of $940$~ns, the sideband relative intensity at $r=305$~nm, the spectral collection range of $\lambda>1.35$~$\upmu$m, and the measured detection losses, we expect $2.5$~kcps from an optimally coupled, unit radiative efficiency $T$ centre. This leaves a factor of $36$ accounted for by some combination of incomplete saturation, uncharacterized losses (e.g. single mode coupling), sub-optimal coupling to the micropuck (e.g. $T$ centre position and orientation) and potentially non-radiative decay processes.

\cref{fig:ple}C shows a scatter plot of the $T$ centre ZPL peak positions and linewidths found within the $144$ micropucks in \cref{fig:ple}A. There is no strong correlation between a centre's linewidth and peak position, even for centres far detuned from the ensemble lineshape. A histogram of the individual peak positions drawn from this scatterplot is shown in \cref{fig:ple}C (top). The distribution of ZPL peak positions is the result of variations in the isotopic and strain environments local to each defect. This inhomogeneous peak distribution is both broader than and slightly shifted from the SOI material ZPL lineshape prior to device patterning from \cref{fig:t_review}C, shown as an overlaid black line. This asymmetry is expected as strain mixes the TX$_0$ and TX$_1$ levels \cite{Safonov1995}.

The observed linewidths match our expectations for single centres in SOI at this temperature. A histogram of the individual linewidths drawn from this scatterplot is shown in \cref{fig:ple}C (right). At saturation power, roughly the power chosen for this study, a typical line will be power broadened from the thermal lower bound (\SI{255}{MHz}) to \SI{361}{MHz} absent spectral diffusion. Including the characteristic spectral diffusion of this material (\SI{1}{GHz}), the typical power broadened linewidth we expect is \SI{1.41}{GHz}, which largely accounts for the median of our measured distribution linewidth, \SI{1.68}{GHz}.

Encouragingly, within nearby $846$~nm radius micropucks we measure individual $T$ centre ZPL lines as narrow as $660$~MHz when driven at low power, from which we conclude that a selection of centres experience less than $400$~MHz total spectral diffusion in this heavily damaged and unoptimized material. As noted in Ref. \cite{MacQuarrie2021}, surface optimization \cite{sangtawesin2019}, electrostatic control \cite{Anderson2019}, and lower levels of implantation damage \cite{vandam2019,Wolfowicz2020a} have all been shown to dramatically reduce environmental noise contributing to spectral diffusion for other colour centres, and similar techniques may be applied to this system.

\paragraph*{Single spins in silicon.}

To further verify that these sparse, resolvable emitters are indeed individual $T$ centres, we observe and characterize the signature spin-selective optical transitions of multiple individually addressable $T$ centre spins in a static magnetic field $B_0$.

The magnetic field $B_0$ splits the ground state isotropic electron spin states $\ket{\uparrow_E}$, $\ket{\downarrow_E}$ as well as the excited state BE anisotropic hole spin states $\ket{\uparrow_H}$, $\ket{\downarrow_H}$. Under these conditions the four spin-selective transitions $A,B,C,D$ shown in \cref{fig:t_review} are no longer degenerate; they are instead detuned from the zero-field ZPL by an amount $\Delta$ according to:
\begin{equation}
    \label{eqn:transitions}
    \Delta_{A,B,C,D} = \dfrac{\mu_B |\vec{B_0}|}{2} \left( \pm g_{E,i} \pm g_{H,i} \right) \,,
\end{equation}
where the $\mu_B$ is the Bohr magneton and $g_{E,i}$,$g_{H,i}$ are the isotropic electron and effective anisotropic hole Land\'{e} $g$ factors for centre $i$ in the specific magnetic field $\vec{B_0}$. Due to the hole $g$ factor anisotropy, each of the twelve orientational subsets of $T$ centres will split differently for a low-symmetry magnetic field direction.

We mount the sample on a permanent SmCo magnet such that the magnetic fringe field magnitude and orientation varies with micropuck position on the chip. For this configuration, we simulate magnetic field magnitudes of $40$~mT~$<|\vec{B_0}|<160$~mT at the device layer of our sample.
Given zero-field ZPL linewidths as low as $660$~MHz this field is sufficient to split and resolve all four spin-selective optical transitions $A,B,C,D$ for some of the twelve $T$ centre orientational subsets. For other orientational subsets, the $B$ and $C$ transitions will not be well resolved even in a magnetic field, because the effective $g$ values for the hole spin and the electron spin for that particular orientation are approximately equal.

As in previous bulk experiments \cite{Bergeron2020}, performing single-frequency PLE spectroscopy in a magnetic field results in almost no PLE signal due to effective electron spin hyperpolarization as the laser scans over the $A,B,C,D$ spin-selective transitions sequentially. The PLE signal which remains is from the subsets of spins whose $B$ and $C$ transitions are not resolved, where no spin shelving state is available and hyperpolarization does not occur. In Ref. \cite{Bergeron2020} the PLE signal from the hyperpolarized (dark) subsets was recovered by depolarizing the spins using magnetic resonance. In this work we depolarize the electron spins by performing two-colour PLE. Specifically, we scan two single-frequency lasers over the individual $T$ centre's ZPL transition energy and observe fluorescence whenever the lasers simultaneously address optical transitions corresponding to the two distinct electron spin states. We filter phonon sideband fluorescence for $\lambda > \SI{1.4}{\um}$ to reject light Raman scattered by the fibre laser combiner used to combine the two single-frequency lasers.

\begin{figure*}[!ht]
    \centering
    \includegraphics[width=17cm]{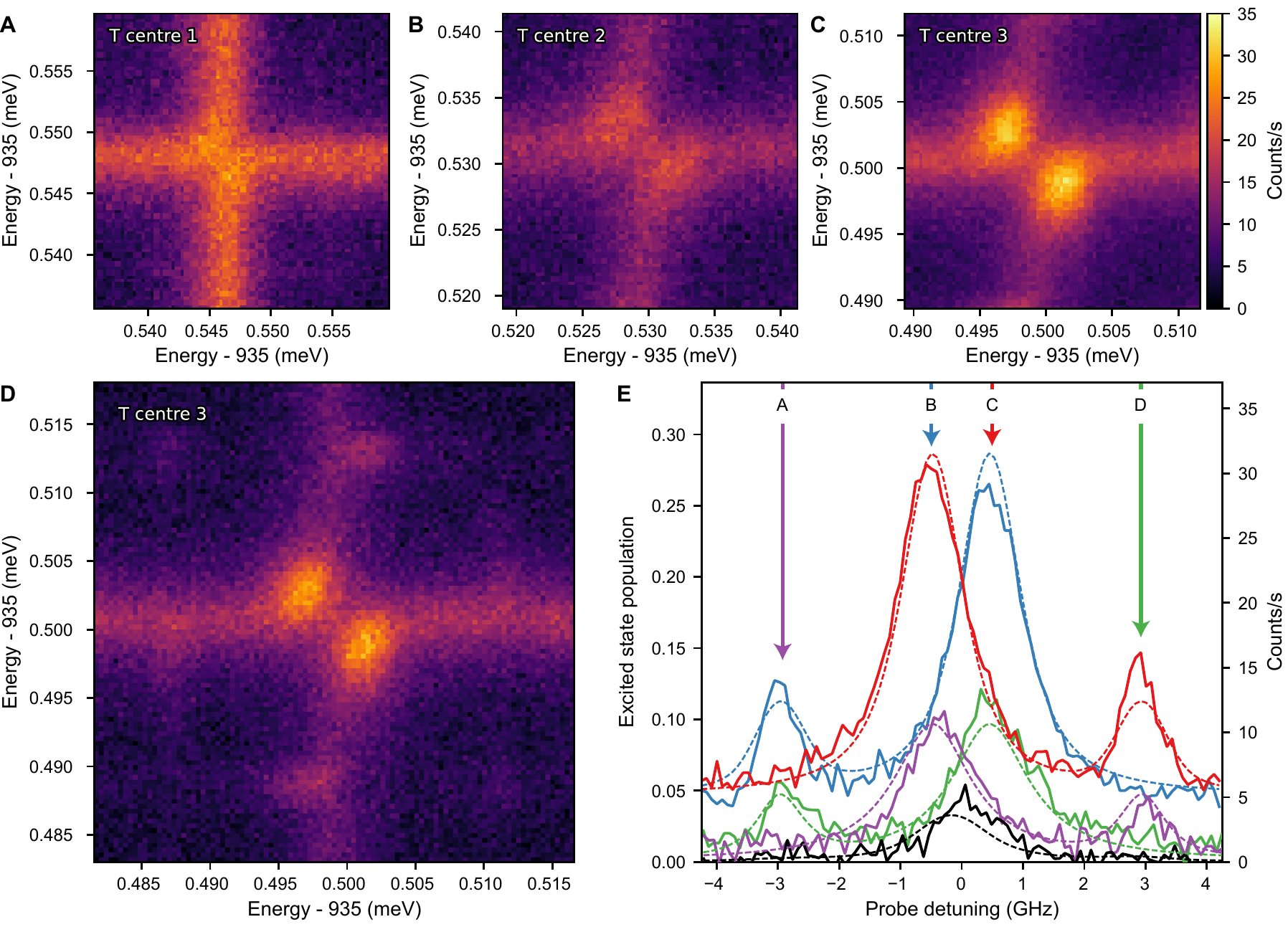}
    \caption{Individual $T$ centre spin-dependent transitions resolved by two-colour resonant excitation in a static magnetic field. (\textbf{A-C}) Two-laser PLE scans over the ZPL of three $T$ centres within different micropucks exhibit varying degrees of splitting. The colour axis is shared. (\textbf{D}) Expanding the resonant excitation range for C reveals four distinct transitions. The colour axis is shared with A-C. (\textbf{E}) Scanning a single probe laser over the resonance produces only weak fluorescence at the overlap of transitions $B$ and $C$ (black). When pumping each of the transitions (labeled), the scanning probe finds brighter resonances (coloured lines) corresponding to transitions from the dark spin state unaddressed by the pump. Pump frequencies are indicated by vertical arrows. Dashed lines indicate a four-level Hamiltonian model fitted simultaneously to the complete set of resonances as discussed in the text.
        \label{fig:spin}
    }
\end{figure*}

In \cref{fig:spin}A-C we show the two-colour PLE frequency scans of three distinct $T$ centres taken from separate $305$~nm radius micropucks shown in \cref{fig:ple}A. These two-colour PLE scans reveal distinct $B$-$C$ transition splittings reflecting the distinct orientations of these $T$ centres relative to the magnetic field.
The centre studied in \cref{fig:spin}A has nearly degenerate $B$ and $C$ transitions, and each laser independently drives continuous fluorescence when addressing these nearly degenerate transitions. The centre studied in \cref{fig:spin}B reveals partially resolved $B$-$C$ transitions, and the centre studied in \cref{fig:spin}C reveals a clear $B$-$C$ splitting of \SI{1}{GHz}. The brightest fluorescence is produced at two-colour combinations where the lasers are detuned from each other and separately resonant with transitions $B$ and $C$.

\cref{fig:spin}D presents an expanded two-colour PLE study of the centre studied in \cref{fig:spin}C, revealing all four resonances $A,B,C,D$. As expected, every combination of laser frequencies which distinctly address the two ground spin states, specifically $A$-$B$, $B$-$C$, $C$-$D$ and $A$-$D$ gives rise to visible PLE signal. We note that, as in Ref. \cite{Bergeron2020}, the labels $A,B,C,D$ refer to specific spin-dependent optical transitions rather than an alphabetical ordering. The effective hole $g$ value relative to that of the electron determines the energy ordering of the $B$ and $C$ transitions.

We perform pump-probe measurements of these two-colour resonances by fixing one `pump' laser at a spin-selective transition and scanning the frequency of the second `probe' laser over the wider range. The coloured lines in \cref{fig:spin}E show probe PLE spectra for a pump at resonances corresponding to each of the transitions $A,B,C,D$ as indicated by a text label and same-colour vertical arrow. For each pump frequency two resonances are visible, corresponding to the two optical transitions connected to the electron spin state unaddressed by the pump. From this data we infer for this $T$ centre the effective hole $g$ factor is larger than that of the electron. A single-frequency PLE scan for reference, in black, shows the weak single-laser resonance at the overlap of the $B$ and $C$ transitions.

Fitting \cref{eqn:transitions} to the peak positions of these spectra gives $|\vec{B_0}| = 88.1(7)$~mT, and $g_\mathrm{H} = 2.76(2)$. We assume $g_\mathrm{E} = 2.005(8)$ for all $T$ centres, as observed in bulk \cite{Bergeron2020}. We note that $|\vec{B_0}|$ is consistent with the calculated fringe field of the permanent magnet at these coordinates. We further fit a four-level optical Bloch Hamiltonian to the pump-probe spectra and find good correspondence to the measured data with a thermal linewidth of \SI{255}{MHz} corresponding to \SI{4.3}{K}, \SI{800(40)}{MHz} Gaussian spectral diffusion and optical Rabi frequencies \SI{10(1)}{MHz}, \SI{ 14(1)}{MHz} for lasers one and two respectively (details in Supplementary Materials). Notably, the observation of a single set of four $A,B,C,D$ transitions for each $T$ centre under study makes the probability that these signals always result from a small plurality of $T$ centres statistically dismissible. For this to occur, the $T$ centres would have to be spectrally identical and oriented identically within the same micropuck in every instance.

\paragraph*{Conclusion.}

We have reported upon the optical detection of individual spins in silicon. We have created tens of thousands of silicon photonic devices, each integrating a small handful of spectrally resolvable and optically accessible $T$ centre spin qubits, leveraging industry-standard photonic SOI wafers and processing. We measure sparse and resolvable $T$ centre optical transitions with linewidths that match the expected linewidths of individual $T$ centre qubits in this material, which are 40 times narrower than the inhomogeneously broadened ensemble linewidth. When placed in a magnetic field, these $T$ centre spin qubits offer spin-selective optical transitions in the telecommunications O-band, and we characterise these spin-dependent optical transitions for a subset of individual $T$ centres using two-colour resonant excitation spectroscopy at temperatures above \SI{4.2}{K}. The anisotropic spin-dependent optical transitions could have in principle revealed up to twelve effective Land\'{e} $g$ factors for small ensembles of $T$ centres; in contrast, for each $T$ centre ZPL studied we extract a single signature effective Land\'{e} $g$ factor, consistent with the interpretation that these are each single centres. Taken together, this work demonstrates that silicon $T$ centres are a suitable technological backbone for commercial-scale, telecommunications-linked, near-term quantum computing and communication networks.

\bibliography{references}
\bibliographystyle{Science}

\section*{Acknowledgments}

The authors thank C. Clément from Polytechnique Montréal for rapid thermal annealing of implanted samples.
\textbf{Funding:} This work made use of the 4D LABS and Silicon Quantum Leap facilities supported by the Canada Foundation for Innovation (CFI), the British Columbia Knowledge Development Fund (BCKDF), Western Economic Diversification Canada (WD) and Simon Fraser University (SFU). This work was supported by the Canada Research Chairs program (CRC), the New Frontiers in Research Fund: Exploration (NFRF-E), the Canadian Institute for Advanced Research (CIFAR) Quantum Information Science program and Catalyst Fund, Le Fonds de recherche du Québec – Nature et technologies (FRQNT), and the Natural Sciences and Engineering Research Council of Canada (NSERC).
\textbf{Author Contributions:} A.T.K.K, D.B.H., and S.S. designed the experiment and wrote the manuscript. A.T.K.K. and D.B.H.  performed the experiment and analysed the data. C.C, D.B.H, E.R.Q., and S.R. developed the samples used in the study. J.R.K., N.R.L-H., J.S., and K.J.M. assisted in experiment design. C.B., L.B., A.D., N.A.B., S.R.H., J.K., M.K., D.W.M., T.S.R., and L.A.S. contributed to code development. M.L.W.T. advised on design and analysis. All authors participated in manuscript revision.
\textbf{Data and materials availability:} Data is available on request.

\section*{Supplementary materials}
Materials and Methods\\
Supplementary Text\\
Figs. S1 to S9\\
Table S1 to S2\\
References \textit{(46-47)}

\clearpage
\setcounter{figure}{0}
\renewcommand{\thefigure}{S\arabic{figure}}
\renewcommand{\thetable}{S\arabic{table}}

\title{Supplementary Materials for \\
    Optical observation of single spins in silicon}

\author
{A. T. K. Kurkjian$^{1\ast}$, D. B. Higginbottom$^{1\ast}$, C. Chartrand$^{1}$,\\
    E. R. MacQuarrie$^{1, 2}$, J. R. Klein$^{1}$, N. R. Lee-Hone$^{1}$, J. Stacho$^{1}$, \\
    C. Bowness$^{1}$, L. Bergeron$^{1}$, A. DeAbreu$^{1}$, N. A. Brunelle$^{1}$, S. R. Harrigan$^{1}$, \\
    J. Kanaganayagam$^{1}$, M. Kazemi$^{1}$,
    D. W. Marsden$^{1}$,  T. S. Richards$^{1}$, L. A. Stott$^{1}$, \\ S. Roorda$^{3}$, K. J. Morse$^{1, 2}$, M. L. W. Thewalt$^{1}$, S. Simmons$^{1,2\dagger}$ \\
    \\
    \normalsize{$^{1}$Department of Physics, Simon Fraser University,}\\
    \normalsize{Burnaby, BC V5A 1S6, Canada}\\
    \normalsize{$^{2}$Photonic  Inc.,  Vancouver,  BC,  Canada}\\
    \normalsize{$^{3}$Département  de physique,  Université  de  Montréal,}\\
    \normalsize{Montréal,  QC H3C 3J7,  Canada}\\
    \normalsize{$^\ast$These authors contributed equally to this work.}\\
    \\
    \normalsize{$^\dagger$Corresponding author. Email: s.simmons@sfu.ca.}
}

\date{}
\pagenumbering{arabic}

\maketitle
\setcounter{page}{0}
\pagenumbering{gobble}
\clearpage
\pagenumbering{arabic}
\setcounter{page}{1}
\section*{Materials and Methods}

\paragraph*{Fourier Transform Infrared Spectroscopy} The photoluminescence (PL) spectra of pre-patterned SOI and bulk \natSi{} samples are measured using a Fourier transform infrared (FTIR) spectrometer. The samples are mounted in a liquid He immersion cryostat and excited above-bandgap using a \SI{532}{\nm} diode laser with a spot size of \SI{3}{\mm}. Fluorescence is analysed by a Bruker IFS 125HR spectrometer to a spectral resolution of \SI{62}{\micro\eV}.

\paragraph*{Confocal Microscopy} The sample is mounted in a low-noise, high optical access cryostat (Montana Instruments s100) with a base temperature of \SI{2.7}{K}. Closed-loop piezoelectric positioners (Attocube, two ANPx101 and one ANPz102) inside the cryostat position the sample along three axes of motion with a \SI{5}{\mm} range and $\approx 1-2$~$\upmu$m precision. An $\text{NA} = 0.71$ microscope objective (SEIWA PE IR 2000HR) is mounted above a fused silica cryostat window. Additional, precise lateral movement is provided by two room temperature piezo positioners (Physik Instrumente P517) that shift the objective by up to $100$~$\upmu$m with $10$~nm precision. The confocal microscope images the sample to a single-mode fibre with a calculated resolution of \SI{1}{\um}.

In the confocal setup PL is generated with a $978$~nm diode laser (QPhotonics QFBGLD-980-5). This above-bandgap light creates free carriers in the silicon which travel and subsequently bind to centres and fluoresce upon recombination. Fluorescence collected by the microscope is directed to either a spectrometer or detector. PL spectra are measured with a fibre-coupled diffraction grating spectrometer (Princeton SpectraPro HRS-300) imaged onto an ultra-low noise LN cooled InGaAs camera (Princeton NIRvana-LN), with a 150 groove/mm grating that provides \SI{1}{\nm} spectral resolution. When using the confocal microscope for imaging or recording photoluminescence excitation (PLE) spectra, fluorescence is directed to a fibre-coupled avalanche photodiode (APD, IDQuantique ID230) configured for 15\% quantum efficiency.

We measure confocal photoluminescence excitation (PLE) spectra using a continuously tunable diode laser (Toptica CTL 1320) locked to a wavemeter (Bristol 871) with $0.01$~pm precision and $1$~pm accuracy. Typical laser power at the Si surface is $\approx 9$~$\upmu$W, or $\approx 7$~$\upmu$W in the Si material accounting for polarization-averaged Fresnel reflection at the surface. The collected PL sideband intensity is measured with the APD. We filter the sideband fluorescence for either $>$\SI{1.35}{\um} (single-colour PLE) or $>$\SI{1.40}{\um} (two-colour PLE).

\section*{Supplementary Text}
\paragraph*{1. Implant and anneal recipe}

Our sample is commercial SOI with a \SI{220}{nm} thick, P-type Czochralski \natSi{} device layer and \SI{3}{\um} thick buried oxide layer. High concentrations of $T$ are created in the device layer by separately implanting carbon-13 and hydrogen, annealing after each implant. Implants were performed by Cutting Edge Ions. Implant energies for the carbon ($38$~keV) and hydrogen ($9$~keV) produce overlapping implant profiles at the device layer centre ($110$~nm depth). Equal doses of $\SI{7e12}{\cm^{-2}}$ are chosen from an implant optimization study reported in \cite{MacQuarrie2021}. Post-implant secondary ion mass spectroscopy (SIMS) measurements confirm carbon is introduced at above the carbon solubility limit. After the initial carbon implant we rapid thermal anneal the sample at \SI{1000}{\celsius} for $20$~s in Argon to repair lattice damage and substitutionalize the implanted carbon. After the hydrogen implant the sample is boiled for $1$~hour in deionized water and, finally, rapid thermal annealed at \SI{420}{\celsius} for three minutes.

\paragraph*{2. Micropuck FDTD simulations}
Three-dimensional finite difference time domain (FDTD) simulations were performed using Lumerical FDTD to compare the collection efficiency $\eta_\text{obj}(\lambda)$ and Purcell factor $P_\text{F}(\lambda)$ from a planar dipole emitter in both an un-etched SOI device layer and silicon micropuck into an $\text{NA} = 0.7$ microscope objective. The relative intensity $I_{\text{r}}(\lambda)$ collected into the objective compared to the total centre fluorescence in a homogeneous medium is
\begin{equation}\label{eqn:irel}
    I_\text{r}(\lambda) = \eta_{\text{obj}}(\lambda) P_\text{F}(\lambda)
\end{equation}

For each of these parameters ($I_\text{r}$, $\eta_\text{obj}$ and $P_\text{F}$) we can derive spectrally weighted integrals corresponding to the intensity, Purcell or efficiency over a useful wavelength range. The weighting function for this integral is the area-normalized bulk $T$ centre spectrum $I_T$ \cite{Bergeron2020}. For example, we may calculate the relative intensity of the $T$ centre phonon sideband:
\begin{align}
    I_\text{r}^\text{PSB} = \int_{\SI{1330}{\nm}}^{\SI{1600}{\nm}} I_\text{r}(\lambda) I_T(\lambda) \dd{\lambda} \,.
\end{align}

\begin{figure*}[!ht]
    \includegraphics[width=16.5cm]{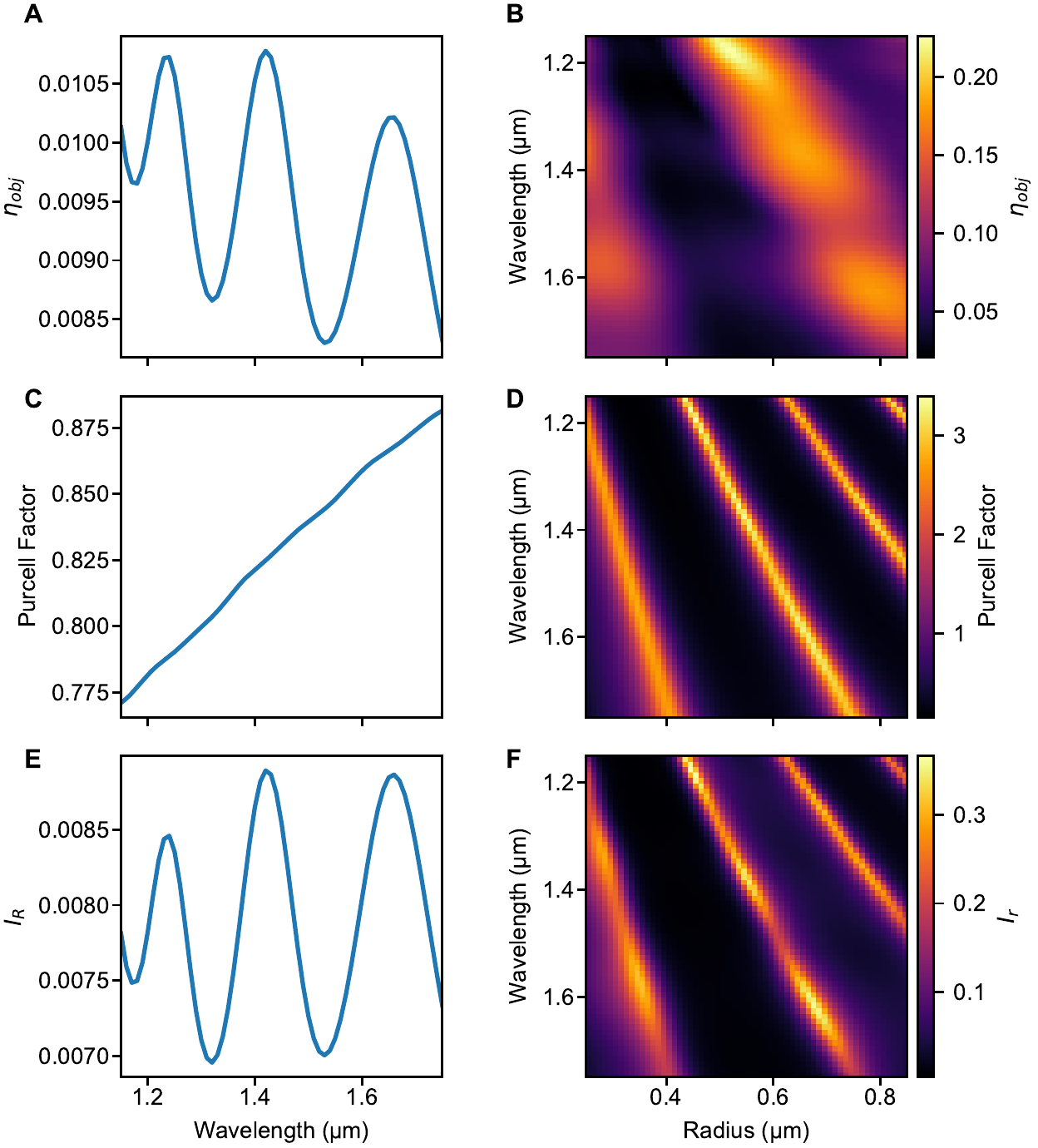}
    \caption{Simulated collection efficiency (\textbf{A} and \textbf{B}), Purcell factor (\textbf{C} and \textbf{D}), and  relative intensity (\textbf{E} and \textbf{F}) for a planar dipole in SOI (left column) and in a micropuck of varying radius (right column).
        \label{fig:sm_pucks}
    }
\end{figure*}

A comparison between the simulated collection efficiency, Purcell factor, and relative intensity in un-etched SOI and the micropucks is provided in \cref{fig:sm_pucks}.

The weighted $I_\mathrm{r}$ for a \SI{305}{nm} micropuck is provided in the table below for the five spectral regions corresponding to the $T$ ZPL, the PL detection range, the PLE detection range,the two-colour PLE detection range, and the total $T$ centre spectrum.
\begin{table}[!ht]
    \begin{center}
        \begin{tabular}{||c c l || c c l ||}
            \hline
            Relative intensity                                   & Range                   & Value \\  [0.5ex]
            \hline\hline
            ZPL, $I^\mathrm{ZPL}_\mathrm{r}$                     & \SI{1.326}{\um}         & 0.021 \\
            PSB (PL detection), $I^\mathrm{PSB}_\mathrm{r}$      & $1.33$--$1.60$~$\upmu$m & 0.153 \\
            PLE detection, $I^\mathrm{PLE}_\mathrm{r}$           & $1.35$--$1.60$~$\upmu$m & 0.146 \\
            Two-colour PLE detection, $I^\mathrm{TC}_\mathrm{r}$ & $1.40$--$1.60$~$\upmu$m & 0.104
            \\
            Full spectrum, $I^T_\mathrm{r}$                      & $1.32$--$1.60$~$\upmu$m & 0.174 \\ [0.5ex]
            \hline
        \end{tabular}
        \caption{Summary of simulated relative intensities.}
        \label{tab:ir}
    \end{center}
\end{table}

\begin{figure*}[!ht]
    \centering
    \includegraphics[width=10cm]{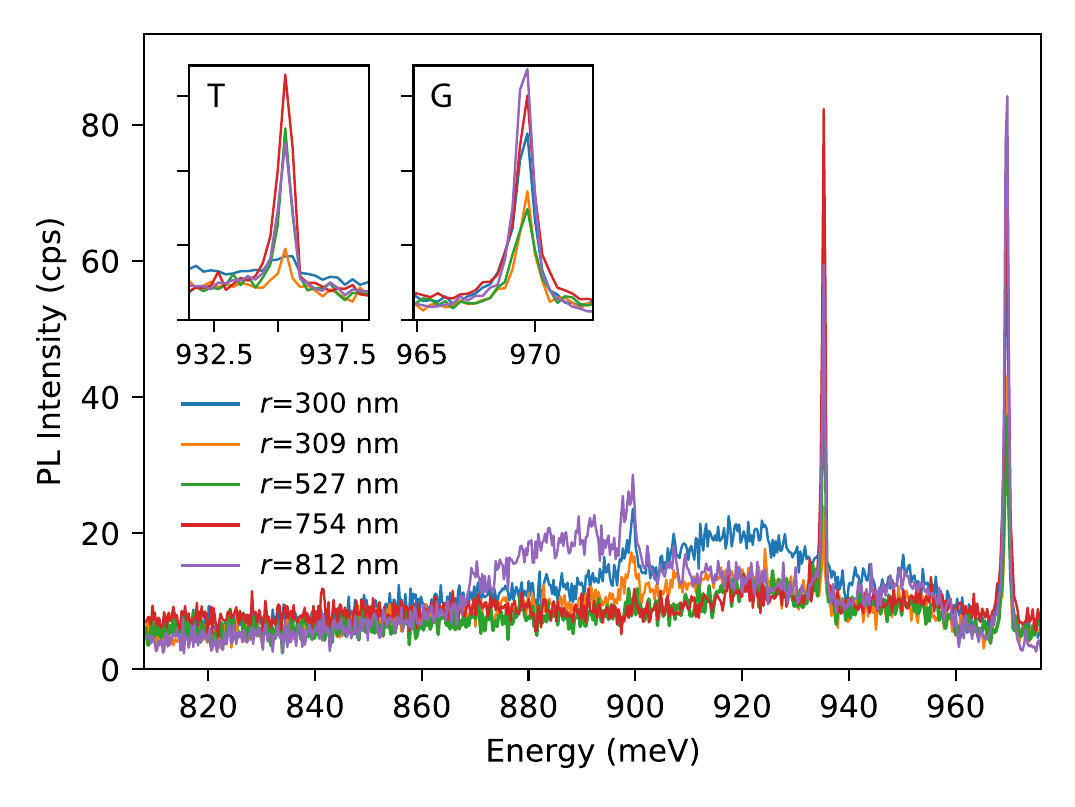}
    \caption{Single micropuck PL spectra.
        \label{fig:puck_spectra}
    }
\end{figure*}

\paragraph*{3. Micropuck PL Spectra}
We measure the PL spectra of micropucks using an ultra-low light spectrometer as described in Materials and Methods. In \cref{fig:puck_spectra} we show spectra for an example selection of individual micropucks of various sizes. The $T$ ZPL evident at \SI{935}{meV} confirms the presence of $T$ in most of these structures. In addition to $T$, we also see the $G$ centre ZPL at \SI{969}{meV}. The $T$, $G$ ZPL peak amplitudes depend sensitively on micropuck radius as the relative intensity enhancement band shifts with radius according to \cref{fig:upucks}B. It can also be seen that particular micropuck radii enhance emission on the $T$ phonon sideband, which is advantageous for PLE measurements. $T$ ZPL areas and integrated phonon sidebands for the complete dataset are shown as a function of radius in \cref{fig:upucks}D.

\paragraph*{4. SOI confocal PLE}

\begin{figure*}[!ht]
    \centering
    \includegraphics[width=8.8cm]{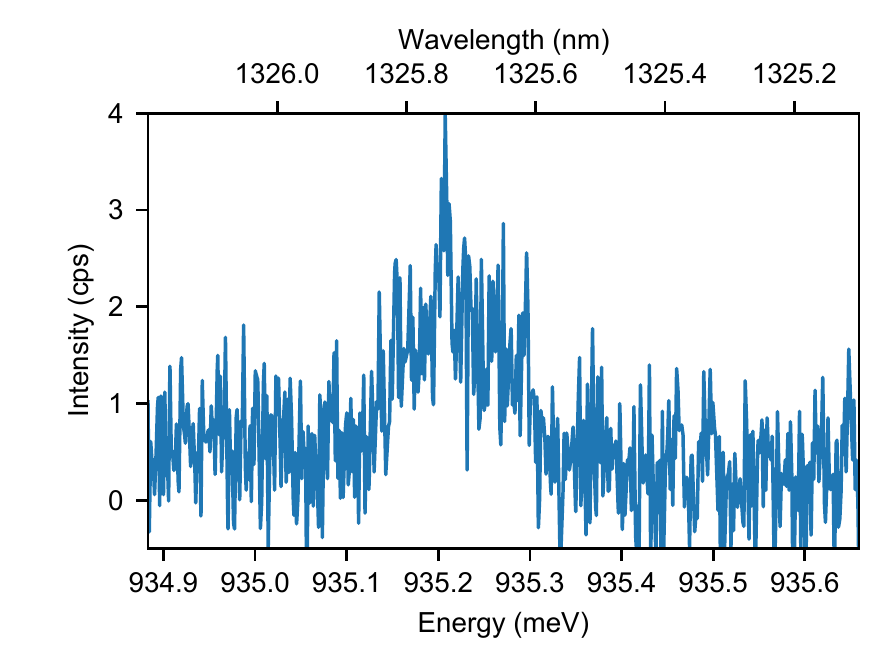}
    \caption{Confocal PLE spectrum of a \SI{200}{\um} large SOI block region. This bulk SOI reference allows us to distinguish the impact of additional interfaces present in  photonic structures like the micropucks.
        \label{fig:soiple}
    }
\end{figure*}

As a reference for the single-centre PLE measurements with micropucks we first measure confocal PLE on a characteristic area within a large block device. The resulting PLE spectrum is shown in \cref{fig:soiple}. We find an inhomogeneously broadened $T$ centre ensemble with a linewidth of \SI{30}{GHz} in this spot, which is somewhat narrower than the \SI{37}{GHz} ZPL linewidth observed from the larger SOI sections in PL both before and after device patterning. The spectrum shows evidence of structure typical of small ensembles on the verge of resolving individual centres. Count rates are very low compared to the micropucks, and confirm the micropuck luminescence advantage expected from simulations.

\paragraph*{5. Low-power PLE}

\begin{figure*}[!ht]
    \centering
    \includegraphics[width=8.8cm]{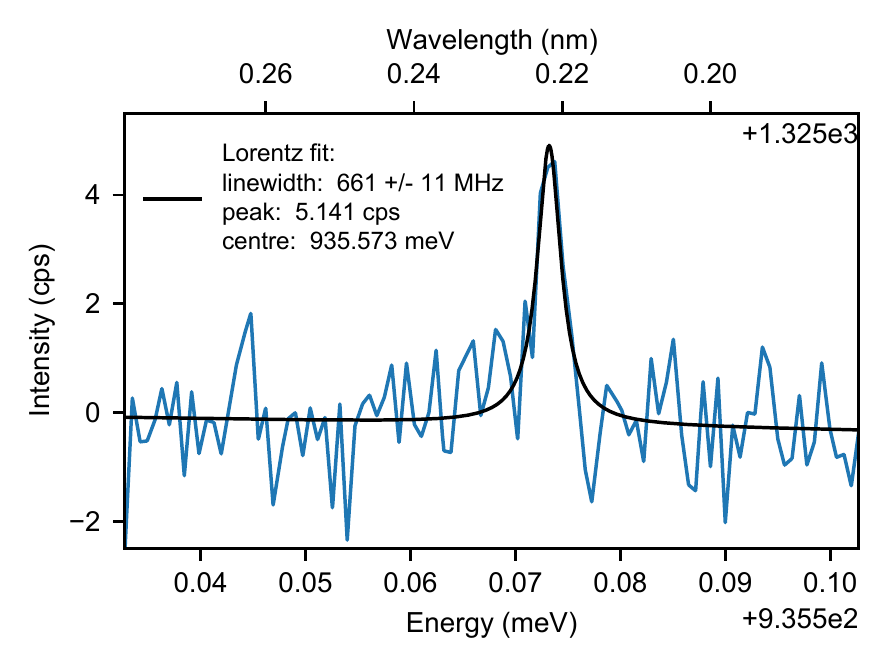}
    \caption{Low-power PLE resonance measured confocally on an \SI{846}{nm} radius micropuck.
        \label{fig:ple_846}
    }
\end{figure*}

The PLE spectra presented in the text were taken over $144$ micropucks at approximately saturation power for a typical centre. Low-power spectra have not been taken systematically in such a large study, but do reveal lower linewidths. The narrowest PLE linewidth we have measured confocally is \SI{660}{MHz}, see \cref{fig:ple_846}, taken on a centre in an {846}{nm} micropuck.

\paragraph*{6. Count rate analysis}

The expected maximum (saturated) count rate from a single $T$ centre is the product of the homogeneous decay rate (halved for resonant excitation), the Purcell factor, and various emission and collection efficiencies, some of which remain unknown. The most significant of these unknowns is the radiative efficiency of the $T$ centre, that is the proportion of decay events that are radiative, which is expected to be near unity \cite{Bergeron2020}. The measured efficiencies relevant to our $305$~nm micropuck PLE measurements are listed below.

\begin{table}[!ht]
    \begin{center}
        \begin{tabular}{||c c l || c c l ||}
            \hline
            Source                                          & Known & Value \\ [0.5ex]
            \hline\hline
            APD, $\eta_\mathrm{D}$                          & spec. & 0.15  \\
            Detection path, $\eta_\mathrm{P}$               & meas. & 0.33  \\
            Cryostat window transmission, $\eta_\mathrm{W}$ & meas. & 0.66
            \\ [1ex]
            \hline
        \end{tabular}
        \caption{Summary of measured losses.}
        \label{tab:eff}
    \end{center}
\end{table}

The micropuck effect captures both the Purcell and the collection efficiency and is given by the relative intensity $I_\mathrm{r}$, with values presented previously in \cref{tab:ir}.

\paragraph*{7. Excited state lifetime}
\begin{figure*}[!ht]
    \centering
    \includegraphics[width=10cm]{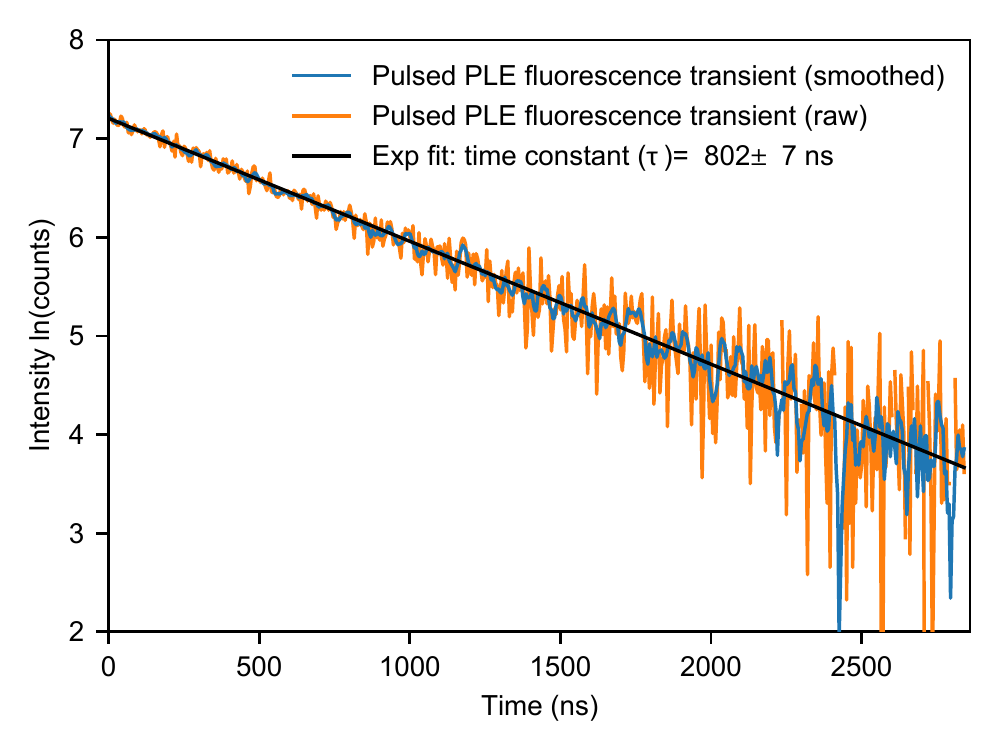}
    \caption{Single $T$ centre excited state lifetime measured by fluorescence transient after resonant excitation An exponential fit gives a lifetime of 802(7) ns.
        \label{fig:ple_lifetime}
    }
\end{figure*}

We measure the excited state lifetime of a micropuck-integrated $T$ centre by pulsed resonant excitation. A single $T$ centre resonance is identified on a $305$~nm micropuck by PLE as shown in \cref{fig:ple}B. The excitation laser is then locked to this frequency and pulsed using an electro-optic amplitude modulator (EOM, Jenoptik AM1310b) with $1$~GHz bandwidth and an extinction ratio of $40$~dB. The excitation laser is pulsed $2$~$\upmu$s on and $3$~$\upmu$s off with a measured optical rise/fall time of $10$~ns. Photon arrival times are tagged using a IDQuantique ID900 time controller.

\begin{figure*}[t!]
    \includegraphics[width=16cm]{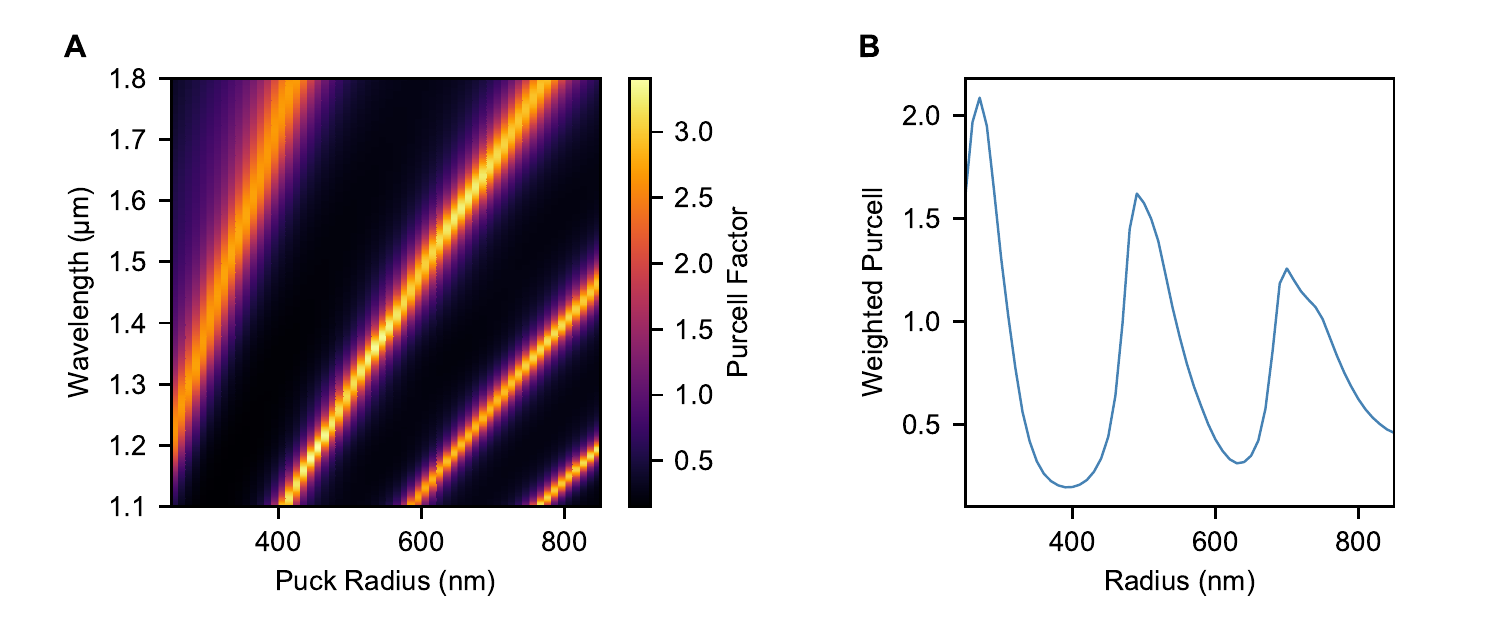}
    \caption{(\textbf{A}) Single-wavelength Purcell factor for a planar dipole at the centre of a micropuck of varying radius. (\textbf{B}) Purcell factor averaged over the $T$ centre spectrum as a function of micropuck radius.}
    \label{fig:purcell}
\end{figure*}

The transient fluorescence decay (shown in \cref{fig:ple_lifetime}) is well fit by an exponential with a time constant of $\tau = 802(7)$~ns, shorter by a factor of $1.17(1)$ than the $940$~ns bulk homogeneous lifetime $\tau_\mathrm{H}$ measured in \TwoEightSi \cite{Bergeron2020}. Purcell enhancement by the micropuck is one mechanism by which the excited-state lifetime may change. \cref{fig:purcell}A shows the Purcell factor $P_\mathrm{f}$ as a function of wavelength and micropuck radius for a single dipole at the micropuck centre oriented in the device plane, simulated using Lumerical FDTD. To find the total Purcell factor for a $T$ centre we take a weighted average of this single-wavelength Purcell over the $T$ spectrum. The total weighted Purcell is shown in \cref{fig:purcell}B. For a $305$~nm micropuck the Purcell factor expected at the micropuck centre is $P_\mathrm{f}^\mathrm{w} = 1.15$.

The lifetime changes according to
\begin{align}
    \frac{\tau}{\tau_\mathrm{H}} = P_\mathrm{f}^\mathrm{w} \eta_\mathrm{R} + (1-\eta_\mathrm{R}) \,,
\end{align}
where $\eta_R$ is the radiative efficiency and every radiative pathway has been included in the average $P_\mathrm{f}$.

The measured lifetime change matches the simulation for unit radiative efficiency $\eta_\mathrm{R}$ but we cannot conclusively attribute the lifetime change to Purcell factor or reach strong conclusions about the radiative efficiency. The exact position and orientation of the centre remain unknown, and alternative explanations for the lifetime change exist. In particular, strain due to the SOI and micropuck interfaces may produce energy eigenstates comprising different momentum states, i.e. mixing the TX$_0$ and TX$_1$ levels.

\paragraph*{8. PLE linewidth analysis}

Our starting point for zero-field PLE linewidth analysis are the Maxwell-Bloch equations for a two level atom in a classical field \cite{Lambropoulos2007}. The steady-state excited state population is
\begin{align}\label{eqn:pee}
    \rho_{ee} = \frac{1}{2} \frac{\Omega ^2}{\left(\frac{\Gamma  \Delta ^2}{\gamma_\mathrm{t} }+\gamma_\mathrm{t}  \Gamma +\Omega ^2\right)} \,,
\end{align}
where $\Omega = \vec{d}\vdot \vec{E}$ is the excitation Rabi frequency, $\Delta$ is the detuning of the field from the transition, $\Gamma$ is the excited state decay rate and $\gamma_\mathrm{t} = \gamma + \Gamma/2$ where $\gamma$ is the transition dephasing rate. At the temperatures studied here $\gamma \gg \Gamma$ and $\gamma_\mathrm{t} \approx \gamma$. $\rho_{ee} < \tfrac{1}{2}$ and approaches $\tfrac{1}{2}$ in the high power limit.
$\rho_{ee}$ is Lorentzian in the field detuning $\Delta$ with FWHM
\begin{align}
    w = 2 \sqrt{\gamma_\mathrm{t}^2 + \Omega^2 \frac{\gamma_\mathrm{t}}{\Gamma}} \,,
\end{align}
where the leftmost part of the sum is the low-power linewidth and the rightmost part of the sum is the linewidth contribution from power broadening.

We define saturation power as $\Omega_\mathrm{s}$ such that $\rho_{ee}(\Omega = \Omega_s, \Delta=0) = \tfrac{1}{2} \mathrm{max}(\rho_{ee}) = \tfrac{1}{4}$, giving $\Omega_\mathrm{s} = \sqrt{\gamma_\mathrm{t} \Gamma}$. The saturation Rabi frequency we expect for $T$ at \SI{4.3}{K} in the absence of spectral diffusion is $\Omega_\mathrm{s} = 4.64\times2\pi$~MHz.

The linewidth at saturation power is therefore
\begin{align}
    w_s = 2 \sqrt{2} \gamma_\mathrm{t} \,,
\end{align}
or $\sqrt{2}$ larger than the zero-power linewidth. From this relationship we arrive at an estimated saturation power linewidth of \SI{361}{MHz} from a thermal dephasing of $\gamma_\mathrm{t} = 127.5\times2\pi$~MHz or a thermal linewidth of \SI{255}{MHz}.

To include spectral diffusion we can follow the same process, but we must first convolve \cref{eqn:pee} by an area-normalized Gaussian spectral diffusion function $G(\Delta)$ with HWHM $\gamma_\mathrm{sd}$ (defined as a HWHM for notational consistency with $\gamma_\mathrm{t}$):
\begin{align}
    \rho_{ee}' = \rho_{ee} * G =  \int_{-\infty}^\infty  \rho_{ee}(\Delta) G(\Delta' - \Delta) \dd{\Delta} \,.
\end{align}
We perform this convolution numerically to find the saturation power $\Omega_\mathrm{s}'$ such that $\rho_{ee}'(\Omega, \Delta'=0) = \tfrac{1}{4}$ and then the FWHM of the spectrally diffused line at $\Omega = \Omega_\mathrm{s}'$ by numerically solving $\rho_{ee}'(\Omega = \Omega_\mathrm{s}', \Delta') = \tfrac{1}{8}$. By this method we arrive at a power-broadened linewidth of \SI{1.41}{GHz} for $\gamma_\mathrm{t} = 127.5\times2\pi$~MHz (\SI{255}{MHz} thermal linewidth) and $\gamma_\mathrm{sd} = 500 \times2\pi$~MHz (\SI{1}{GHz} spectral diffusion). The corresponding saturation Rabi frequency is $\Omega_\mathrm{s}' = 11.4\times2\pi$~MHz.

\paragraph*{9. Four-level spin-optical Hamiltonian}

We fit a four-level, semiclassical atom-light model to pump-probe measurements of the $\vec{B}$ field-split $T$ PLE spectrum. For convenience of notation we label the spin states and transitions in order of energy (when $g_\mathrm{H} > g_\mathrm{E}$ as it is here): $\ket{\downarrow_\mathrm{E}}$, $\ket{\uparrow_\mathrm{E}}$, $\ket{\downarrow_\mathrm{H}}$, $\ket{\uparrow_\mathrm{H}}$ $\rightarrow$ $\ket{1}$, $\ket{2}$, $\ket{3}$, $\ket{4}$ and $A$, $B$, $C$, $D$ $\rightarrow$ 1,2,3,4. We solve the evolution of the reduced atomic density matrix $\qo{\rho}$ under excitation by either one or two classical fields according to the master equation
\cite{Lambropoulos2007}
\begin{equation}\label{eqn:master}
    \pdv{t} \qo{\rho}  = -\frac{i}{\hbar} \comm{\qo{H}}{\qo{\rho}} + \qo{\mathcal{L}}(\qo{\rho}) \,.
\end{equation}

We separate the Hamiltonian $\qo{H}$ into atomic and interaction components in order to work in the interaction picture. We further make the rotating wave approximation to reduce the electric dipole interaction Hamiltonian to
\begin{align}
    \hat{H}_\mathrm{i} = -\frac{\hbar}{2} \sum_{i,j} \Omega_{ij} \hat{\sigma}_{ij} + \mathrm{h.c} \,,
\end{align}
where $\sigma_{ij} = \dyad{i}{j}$ is the atomic raising operator from state $\ket{i}$ to $\ket{j}$ and $\Omega_{ij} = \vec{d}_{ij} \vdot \vec{E_{ij}}$ is the corresponding Rabi frequency, determined by the transition dipole moment $\vec{d}_{ij}$ and resonant electric field amplitude $\vec{E}_{ij}$.
The atomic Hamiltonian is
\begin{align}
    \hat{H}_\mathrm{a} & = -\frac{\hbar}{2}\left(\Delta^\mathrm{L}_1 \qsigma_{11} + \Delta^\mathrm{L}_2 \qsigma_{22} +  \sum_{i=1}^4 \Delta_{i}^\mathrm{Z} \hat{\sigma}_{ii} \right) \\
    \Delta^{Z}_{i}     & = \dfrac{\mu_B |\vec{B_0}|}{2} \left( \pm g_{E} \pm g_{H} \right) \,,
\end{align}
where $\Delta_i^\mathrm{Z}$ is the Zeeman splitting of level $i$ (\cref{eqn:transitions} in the main text) and $\Delta_i^\mathrm{L}$ is the detuning of the laser addressing ground states $i=1,2$ from the zero-field transition energy.
The non-unitary evolution of the system is captured through a Liouvillian superoperator containing all dissipative system dynamics. The Liouvillian has Lindblad form
\begin{equation}\label{eqn:liouvillian}
    \qo{\mathcal{L}} = \sum \tfrac{1}{2} \left( 2 \qo{C} \qo{\rho} \dg{\qo{C}} - \dg{\qo{C}} \qo{C} \qo{\rho} - \qo{\rho} \dg{\qo{C}} \qo{C} \right) \,,
\end{equation}
where $\sum$ is a sum over the following collapse operators $\qo{C}$. Spontaneous decay (radiative or otherwise) is modelled with the atomic transition collapse operator $\qo{C}^\mathrm{r}_\mathrm{ij} = \sqrt{\Gamma_{ij}} \qo{\sigma}_{ij}$ where $\Gamma_{ij}$ is the transition decay rate. Optical dephasing is included by adding additional collapse operators $\qo{C}^\mathrm{d}_\mathrm{ij} = \sqrt{\frac{\gamma}{2}}\left( \qsigma_{ii} - \qo{\sigma}_{jj} \right)$ where $\gamma$ is a constant dephasing rate and the low-power linewidth is $(\Gamma + 2 \gamma)/(2\pi)$~GHz. For each of these operators $i=3,4$ and $j=1,2$.

We solve the master equation evolution using QuTiP, the quantum toolbox package for Python \cite{Johansson2012}. The fluorescence rate is modeled by the steady-state excited population $\expval{\qsigma_{22}} + \expval{\qsigma_{33}}$ of \cref{eqn:master}. The excited state population data used for fitting is determined from the detected fluorescence by normalization to a saturated fluorescence value. The five pump-probe datasets in \cref{fig:spin}E are fit simultaneously with a single model. The excited state decay rate  and electron spin Land\'{e} factor  are fixed to their bulk Si values \cite{Bergeron2020}, $\Gamma = 1/\tau_\mathrm{H} = 169\times2\pi$~MHz and $g_\mathrm{E} = 2.005$. This model does not account for lifetime changes induced by either Purcell or strain effects. $\gamma$ is fixed to the thermal dephasing at rate at \SI{4.3}{K}, $\gamma = 255/2 \times 2 \pi = 127.5 \times 2 \pi$~MHz.

Once again, spectral diffusion is incorporated by convolution by the normalized Gaussian spectral diffusion function $G(\Delta')$. We integrate over $H_\mathrm{a}$ numerically.
\begin{align}
    \hat{H}_\mathrm{a} & = -\frac{\hbar}{2} \int G(\Delta') \left( \left(\Delta^\mathrm{L}_1 - \Delta' \right) \qsigma_{11} + \left(\Delta^\mathrm{L}_2 - \Delta' \right) \qsigma_{22}\right)\dd{\Delta'} +  \sum_{i=1}^4 \Delta_{i}^\mathrm{Z} \hat{\sigma}_{ii} \,.
\end{align}
The SD HWHM $\gamma_\mathrm{sd}$, $\vec{B}_0$ and $g_\mathrm{H}$ are free fit parameters. We further define free fit parameters $\Omega_{\mathrm{p,r}}$ for each of the probe and repump excitation laser powers (expressed as Rabi frequencies).

The transition Rabi frequencies  $\Omega_{ij}$ and laser detunings $\Delta^\mathrm{L}_{1,2}$ are determined separately for each pump-probe dataset according to the scan configuration. One of $\Delta^\mathrm{L}_{1,2}$ is varied according to the probe scan range and the other is fixed at the pump value. Rabi frequencies $\Omega_{i,j}$ are set correspondingly to one of the laser powers $\Omega_\mathrm{p}$, $\Omega_\mathrm{r}$. depending on which laser is fixed and which is being scanned. The single-frequency PLE dataset is simulated similarly using a single field addressing both transitions $B$ and $C$.

The transition decay rates $\Gamma_{i,j}$ are scaled by a free branching ratio parameter $r$ that describes the relative dipole strength of the spin preserving transitions B,C to the spin changing transitions A,D.
\begin{align}
    \Gamma_{3,1} = \Gamma_{4,2} & = \frac{r}{1+r} \Gamma     \\
    \Gamma_{4,1} = \Gamma_{3,2} & = \frac{1}{1+r} \Gamma \,.
\end{align}
The transition Rabi frequencies are $\Omega_{i,j}$ scaled by $r$ and also a free relative transition dipole/polarization overlap parameter $p$ assuming that transitions $A$,$D$ ($B$,$C$) have the same dipole orientation. We fit $r=1.2(2)$ and $p=2.9(3)$ along with the values given in the text.

With a free dephasing rate $\gamma$, the fit process returns $\gamma = 185(20)\times2\pi$  (\SI{370}{MHz} FWHM), $\gamma_\mathrm{sd} = 230(80) \times 2\pi$~MHz (\SI{460}{MHz} FWHM), $\Omega_1 = 9(1)\times2\pi$~MHz, $\Omega_2 = 11(1)\times2\pi$~MHz, $r = 1.6(2)$, $p=2.6(2)$. This implies a temperature of \SI{4.7}{K}, which is slightly above the \SI{4.3(3)}{K} uncertainty range of our TX$_0$/TX$_1$ PL ratio temperature estimate. It is possible that any given centre has more or less dephasing than this model accounts for due to strain induced shifts of the TX$_0$ and TX$_1$ levels. The dephasing rate at a given temperature changes inversely with the TX$_0$--TX$_1$ splitting. The free fit compensates for this increased dephasing rate by decreasing the spectral diffusion to \SI{460}{MHz}, close to the lower bound of what we've observed to date. In practise this exchanges a Gaussian linewidth contribution for a Lorentzian linewidth contribution, and may reflect that the spectral diffusion process is not strictly Gaussian.

\clearpage

\end{document}